\documentstyle[11pt,newpasp,twoside,epsf]{article}
\def\edcomment#1{\iffalse\marginpar{\raggedright\sl#1\/}\else\relax\fi}
\reversemarginpar

\begin{document}
\title{The Formation of the Most Relativistic Pulsar PSR J0737-3039}
\author{B. Willems, V. Kalogera, and M. Henninger}
\affil{Northwestern University, Department of Physics and Astronomy,
    2145 Sheridan Road, Evanston, IL 60208, USA}

\begin{abstract}
We present an updated set of constraints for the progenitor of
PSR\,J0737-3039 and for the natal kicks imparted to pulsar~B taking
into account both the evolutionary and kinematic history of the double
neutron star. For this purpose, we use recently reported scintillation
velocity measurements to trace the motion of the system in the Galaxy
backwards in time as a function of the unknown orientation $\Omega$ of
the systemic velocity projected on plane of the sky as well as the
unknown radial velocity $V_r$. The absolute limits on the orbital
separation and the mass of pulsar~B's helium star progenitor just
before its supernova explosion are $1.2\,R_\odot \la A_0 \la
1.7\,R_\odot$ and $2.1\,M_\odot \la M_0 \la 4.7\,M_\odot$. The kick
velocity is constrained to be between 60\,km\,s$^{-1}$ and
1660\,km\,s$^{-1}$ and to be misaligned from the pre-SN orbital
angular axis (which could be associated with pulsar~B's spin axis) by
least $25^\circ$ . We also derive probability distribution functions
for the kick velocity imparted to pulsar~B and for the misalignment
angle between pulsar~A's spin and the post-supernova orbital angular
momentum for both isotropic and polar kicks. The most probable values
of both quantities depend sensitively on the unknown radial
velocity. In particular, tilt angles lower than $30^\circ$--$50^\circ$
tend to be favored for current radial velocities of less than $\simeq
500$\,km\,s$^{-1}$ in absolute value, while tilt angles higher than
$120^\circ$ tend to be associated with radial velocities in excess of
$\simeq 1000$\,km\,s$^{-1}$ in absolute value.
\end{abstract}

\section{Introduction}

The recent discovery of the strongly relativistic binary pulsar
(Burgay et al. 2003) which is also the first eclipsing double pulsar
system found in our Galaxy (Lyne et al. 2004) has resparked the
interest in the evolutionary history and formation of double neutron
star (DNS) systems (Willems \& Kalogera 2004; Dewi \& van den Heuvel
2004; Willems, Kalogera \& Henninger 2004). According to the standard
formation channel for DNS binaries (e.g. Bhattacharya \& van den
Heuvel 1991), their progenitors evolve through a high-mass X-ray
binary phase where the first-formed NS accretes matter from the wind
of a high-mass companion. When the latter fills its Roche lobe, the
extreme mass ratio triggers the formation of a common envelope which
extracts orbital energy and angular momentum and causes the NS and
donor core to spiral-in towards each other. If the inspiral can be
stopped before the components coalesce, a tight binary is formed
consisting of the first-formed NS and a helium star companion. After
it exhausts its central helium supply, the helium star expands and a
second, this time stable, mass-transfer phase is initiated which spins
the NS up to millisecond periods. At the end of the mass-transfer
phase the helium star's core explodes and forms the second NS.

In this paper, we present constraints on the properties of the
progenitor of PSR\,J0737-3039 just before the supernova (SN) explosion
that gives birth to pulsar~B. In view of the recent scintillation
velocity measurements, which were first presented at the meeting
(Ransom et al., these proceedings) and updated shortly thereafter
(Ransom et al. 2004), we update our results presented at the
conference and account for the system's full kinematic history since
the time of its formation. We furthermore present probability
distribution functions (PDFs) for the kick velocity imparted to
pulsar~B and, in reply to the call for theorists' predictions on
PSR\,J0737-3039, we predict the most likely misalignment angle between 
pulsar~A's spin and the post-SN orbital angular momentum as a function
of $V_r$.

\section{Progenitor constraints}

In order to derive constraints on the pre-SN binary parameters and the
kick imparted to pulsar~B, we first use the scintillation velocity
components determined by Ransom et al. (2004) to trace the Galactic
motion of the system backwards in time as a function of the unknown
orientation $\Omega$ of the scintillation velocity in the plane of the
sky and the unknown radial velocity $V_r$. Assuming that the system is
at most 100\,Myr old, we find that it may have crossed the Galactic
plane twice in the past. Identifying these plane crossings with
possible birth sites yields estimates for the age and post-SN peculiar
velocity which are shown in Fig.~1 as a function of $\Omega$ and
$V_r$. If the system has crossed the Galactic plane only once in the
past, there is a wide range of $\Omega$- and $V_r$-values for which
PSR\,J0737-3039 may be remarkably young ($\la 20$\,Myr). If, on the
other hand, the system has crossed the Galactic plane twice, it is at
least 20\,Myr old regardless of the values of $\Omega$ and $V_r$. The
peculiar velocity right after the formation of pulsar~B may
furthermore be anywhere between $\simeq 90$\,km\,s$^{-1}$ and $\simeq
1200$\,km\,s$^{-1}$.

\begin{figure}
\plottwo{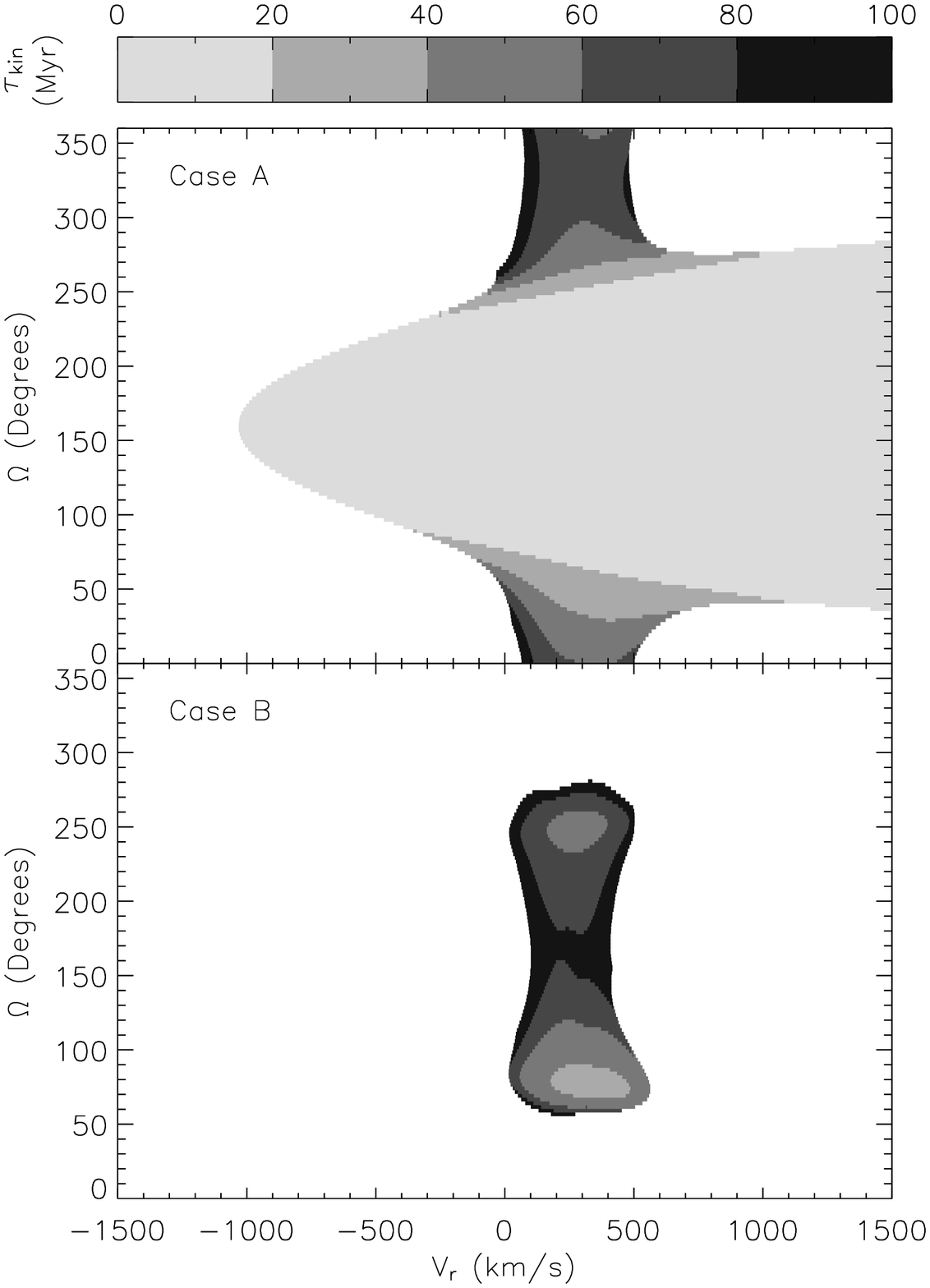}{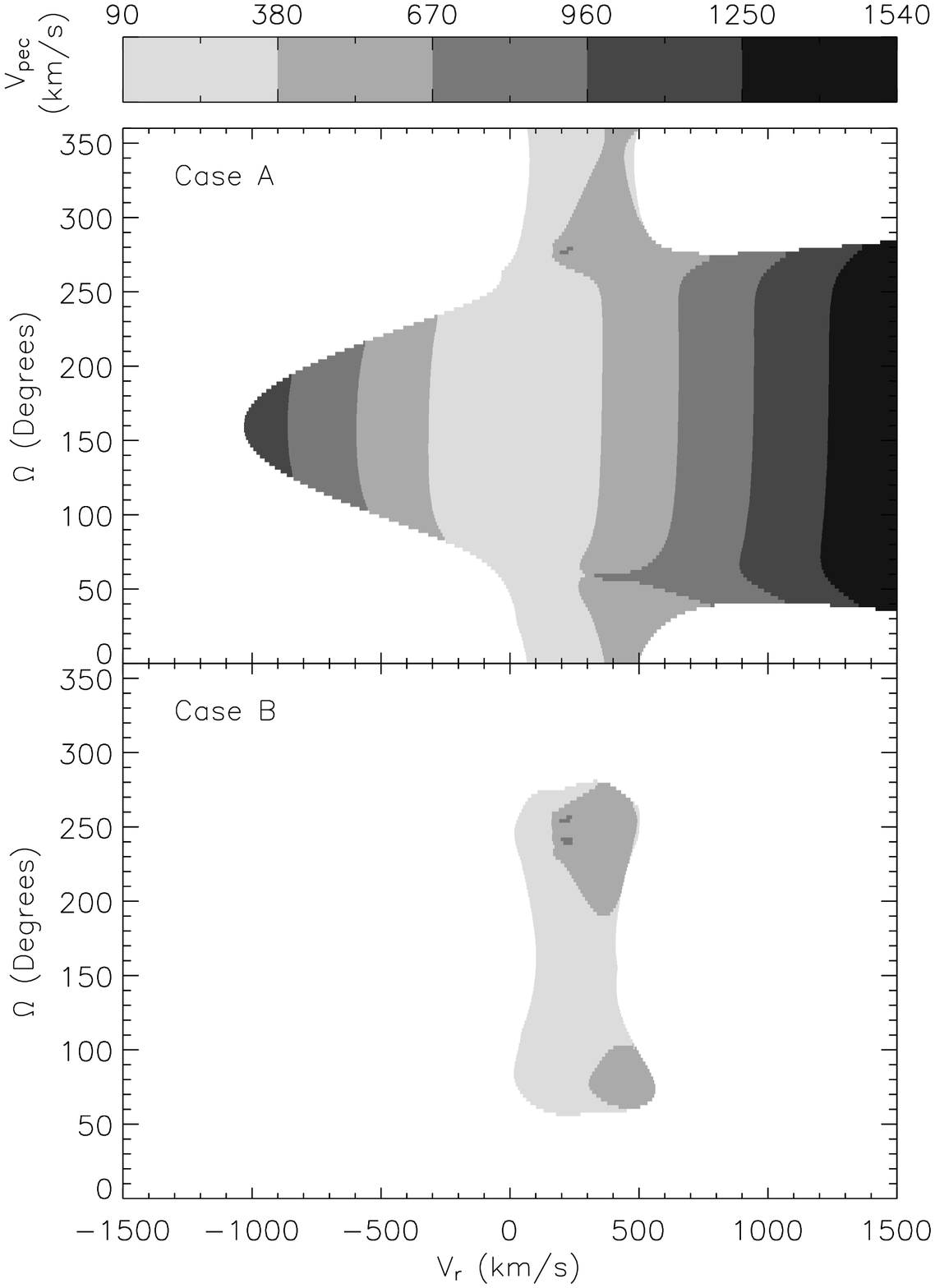}
\caption{Kinematic age $\tau_{\rm kin}$ (left) and post-SN peculiar
  velocity $V_{\rm pec}$ (right) of PSR\,J0737-3039 as a function of
  $\Omega$ and $V_r$. Cases~A and~B respectively refer to the first
  and second Galactic plane crossing when following the orbit of
  the system backwards in time.}
\end{figure}

Next, we determine the post-SN orbital parameters for each possible
birth site by integrating the equations governing the orbital
evolution due to gravitational radiation backwards in time up to the
age corresponding to that birth site. These post-SN binary parameters
are related to the pre-SN parameters and to the kick
magnitude and direction by the conservation laws of orbital energy and
angular momentum. The requirements that (i) the solutions to these
relations must be real, (ii) the binary components must remain bound
after the SN explosion, (iii) the post-SN orbit must pass through the
position of both stars at the time of the SN explosion, and (iv) the
kick must produce the right post-SN peculiar velocity, then impose
constraints on the pre-SN orbital separation and on the mass of
pulsar~B's helium star progenitor as well as on the magnitude and
direction of the kick imparted to pulsar~B (for more details, see
Willems et al. 2004, and references therein).

Regardless of the values of the two unknown parameters $\Omega$ and
$V_r$, the pre-SN orbital separation is constrained to the interval
between $1.2\,R_\odot$ and $1.7\,R_\odot$. For these tight
constraints, the system is able to avoid Roche-lobe overflow prior to
the formation of pulsar~B only if the mass of B's helium star
progenitor is higher than $25\,M_\odot$ and if the kick velocity
imparted to pulsar~B at birth was larger than $\simeq
1200$\,km\,s$^{-1}$. Helium stars of such a high mass are not only
very unlikely (due to the strong wind mass loss), they are also
expected to end up as a black hole instead of a NS. The pre-SN binary
therefore most likely consisted of the first-formed NS in orbit around
a Roche-lobe filling helium star (see also Dewi \& van den Heuvel
2004). We constrain the mass of the latter to be between
$2.1\,M_\odot$ and $4.7\,M_\odot$, where the lower mass limit
corresponds to the smallest mass for which a single helium star is
expected to form a NS and the upper limit to the highest mass for
which mass transfer onto the first-formed NS is expected to be
dynamically stable (Ivanova et al. 2003). The lower limit of
$2.1\,M_\odot$ furthermore implies that a birth kick with a velocity
of at least 60\,km\,s$^{-1}$ was imparted to the second-born NS, in
agreement with the minimum kick velocity derived by Willems \&
Kalogera (2004) and Dewi \& van den Heuvel (2004). From the condition
that the binary must remain bound after the second SN explosion, we
also derived an upper limit for the kick velocity of
1660\,km\,s$^{-1}$. The direction of the kick is constrained to always
make an angle of at least $115^\circ$ with the helium star's pre-SN
orbital velocity, and and angle of at least $25^\circ$--$30^\circ$
with the pre-SN orbital angular momentum axis (which could be
associated with pulsar B's spin axis). Hence, the kicks are generally
directed opposite to the orbital motion and could not have been too
closely aligned or anti-aligned with the pre-SN orbital angular
momentum.

For specific values of $\Omega$ and $V_r$, the progenitor and kick
velocity properties often become much more constrained than the
absolute limits mentioned above. In particular, the condition that the
kick must match the post-SN peculiar velocity associated with the
considered values of $\Omega$ and $V_r$ narrows the range of possible
kick velocities to an interval that is about $\simeq
500$\,km\,s$^{-1}$ wide. The minimum kick velocity and the tendency of
the kick to be directed opposite to the pre-SN orbital motion
furthermore increase with increasing absolute values of the radial
velocity. For a detailed representation of the progenitor and kick
constraints as a function of $\Omega$ and $V_r$, we refer to Willems
et al. (2004). 

\section{Isotropic kick-velocity and spin-tilt distributions}

Following the procedure described in detail in Willems et al. (2004),
we derive probability distribution functions (PDFs) for the kick
velocity $V_k$ imparted to pulsar~B at birth and for the misalignment
angle $\lambda$ between pulsar~A's spin and the post-SN orbital
angular momentum under the assumption that all kick directions are
equally probable. The PDFs are first derived for all admissible values
of $\Omega$ and $V_r$, and next integrated over all possible
$\Omega$-angles assuming a uniform distribution between 0 and
$2\pi$. The resulting PDFs are shown in Figs.~2 and~3 as functions of
the remaining unknown parameter $V_r$.

\begin{figure}
\plotone{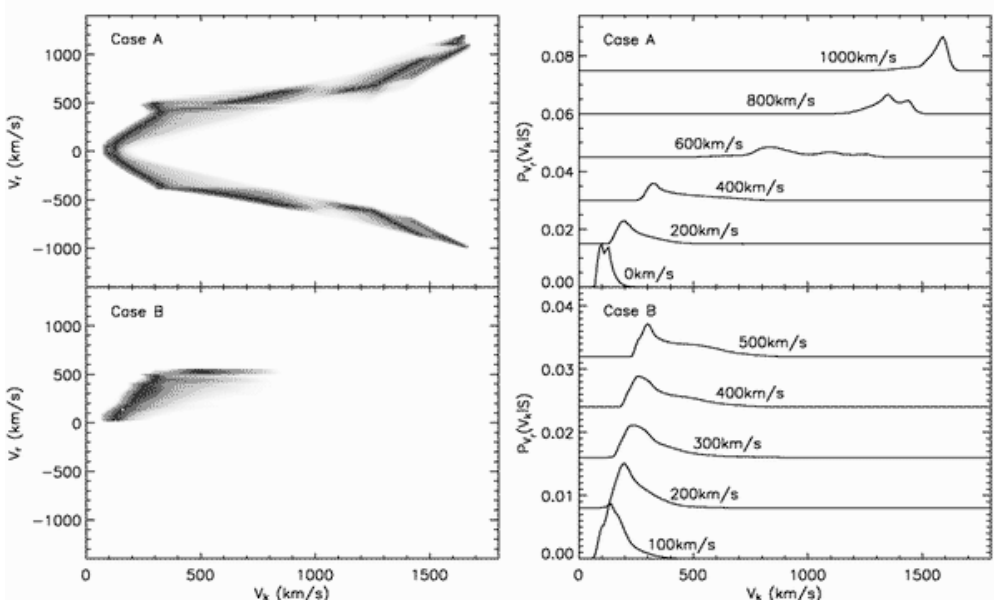}
\caption{Kick probability distribution functions for
PSR\,J0737-3039B. Left panel: PDFs associated with all admissible
$V_r$-values. The gray scale is linear and varies from light to dark
gray with increasing PDF-values. Right panel: PDFs associated with some
specific $V_r$-values (offset by an arbitrary amount).}
\end{figure}

\begin{figure}
\plotone{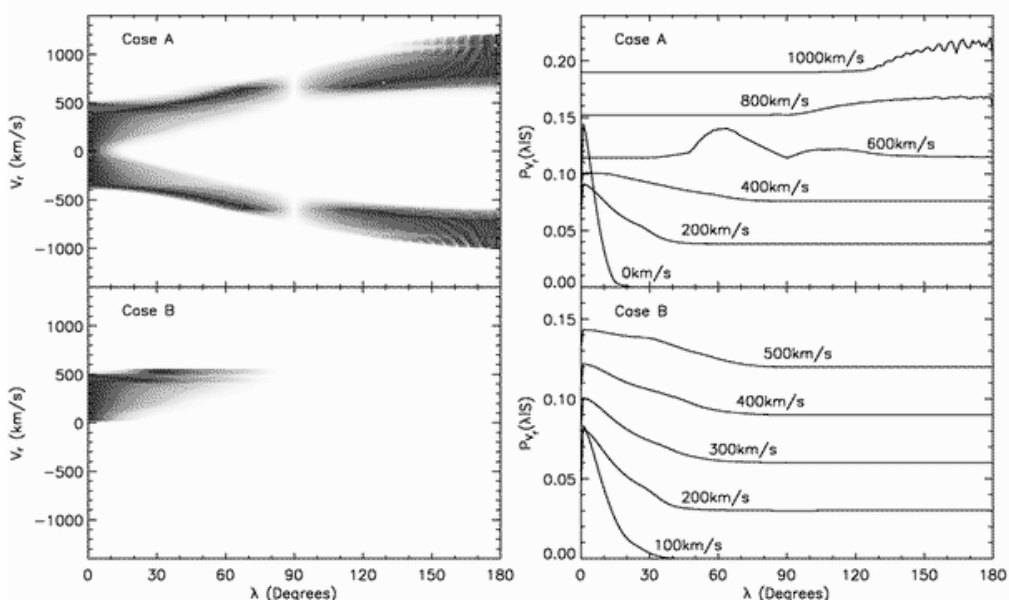}
\caption{As Fig.~1, but for the spin-orbit misalignment angle
  $\lambda$.}
\end{figure}

The kick-velocity distributions exhibit a single peak at kick
velocities which increase from $\simeq 100$\,km\,s$^{-1}$ to $\simeq
1600$--1700\,km\,s$^{-1}$ with increasing absolute values of the
radial velocity. This is in contrast to the single most probably kick
velocity of $\simeq 150$\,km\,s$^{-1}$ found by Willems \& Kalogera
(2004). The wider range found here is associated with the relation
between the kick magnitude and the post-SN peculiar velocity which
tends to select those kick velocities from the admissible range which
are able to explain the motion of the system in the Galaxy (high space
velocities cannot be explained by low kick velocities and vice
versa). However, for a given radial velocity, the range of kick
velocities with non-vanishingly small probabilities is always much
more constrained than in Willems \& Kalogera (2004).

The tilt-angle distributions, on the other hand, clearly favor tilt
angles below $30^\circ$--$50^\circ$ when (i) the system has crossed
the Galactic plane twice in the past or (ii) the system has crossed
the Galactic plane once and has a current radial velocity of less than
$\simeq 500$\,km\,s$^{-1}$ in absolute value. Higher tilt angles tend
to be associated with one disk crossing and much higher radial
velocities of $\simeq 500$--1200\,km\,s$^{-1}$ in absolute value. Tilt
angles close to $90^\circ$ are furthermore strongly disfavored for any
radial velocity $V_r$.  The overall behavior of the PDFs thus supports
the tilt angles of $16^\circ \pm 10^\circ$ or $164^\circ \pm 10^\circ$
predicted by the geometrical model of Jenet \& Ransom (2004), as well
as the problems with their alternative solutions of $82^\circ \pm
16^\circ$ or $98^\circ \pm 16^\circ$ which are incompatible with the
misalignment angle between the pulsar's spin and magnetic dipole axes
derived by Demorest et al. (2004).

\section{Non-isotropic kicks}

In view of recent claims of alignment of NS kicks with NS spin axes
(e.g. Romani 2004, these proceedings), it is also interesting to
consider the constraints on the progenitor and formation of
PSR\,J0737-3039 in the case where the kick direction is confined
within two oppositely directed cones with an opening angle of
$30^\circ$ and with axes parallel to the pre-SN orbital angular
momentum axis (i.e. polar kicks). The confinement of the kick
directions strongly reduces the range of radial velocities for which
viable progenitors for PSR\,J0737-3039 may be found and greatly
tightens the constraints described in the previous sections: the mass
of pulsar~B's helium star progenitor becomes constrained to $M_0
\simeq 2.1$--$2.5\,M_\odot$, the pre-SN orbital separation to $A_0
\simeq 1.1$--$1.5\,R_\odot$, and the magnitude of the kick velocity
to $V_k \simeq 100$--600\,km\,s$^{-1}$. The range of most probable
kick velocities furthermore shrinks to $\simeq
200$--550\,km\,s$^{-1}$, and the range of most probable tilt angles to
$15^\circ$--$45^\circ$. 

\section{Concluding remarks}

We have used the scintillation velocity measurements of Ransom et
al. (2004) to derive the most up-to-date constraints on the progenitor
of PSR\,J0737-3039 and on the kick imparted to pulsar~B at birth as a
function of the unknown orientation $\Omega$ of the systemic velocity
projected on plane of the sky and the unknown radial velocity
$V_r$. We also derived the most probable pulsar kick velocity and spin
tilt for both isotropic and polar kicks as a function of $V_r$. Once
$\Omega$ is measured in the coming year, it will be straightforward to
use the results presented here to further constrain the natal kicks
and the spin-tilt predictions. 

\acknowledgments

We thank the Aspen Center for Physics and the conference organizers
for hosting a very stimulating meeting, and Laura Blecha for sharing 
the Galactic motion code. This work is partially supported by a NSF
Gravitational Physics grant, a David and Lucile Packard Foundation
Fellowship in Science and Engineering grant, and NASA ATP grant
NAG5-13236 to VK.


\begin{references}
\reference Bhattacharya, D. \& van den Heuvel, E.P.J.\ 1991,
Phys. Rep., 203, 1 
\reference Burgay, M., et al. 2003, Nature, 426, 531
\reference Demorest, P., Ramachandran, R., Backer, D.C., Ransom,
  S.M., Kaspi, V., Arons, J., \& Spitkovsky, A.\ 2004, ApJL, submitted
  (astro-ph/0402025) 
\reference Dewi, J.D.M. \& van den Heuvel, E.P.J. 2004, MNRAS, 349,
169 
\reference Ivanova, N., Belczynski, K., Kalogera, V., Rasio, F.A. 
  \& Taam, R.E. 2003, ApJ, 592, 475
\reference Jenet, F.A. \& Ransom, S.M. 2004, Nature, in press
\reference Lyne, A.G., et al. 2004, Science, 303, 1153 
\reference Ransom, S.M., et al. 2004, ApJL, submitted (astro-ph/0404149)
\reference Willems, B. \& Kalogera, V. 2004, ApJL, 603, L101
\reference Willems, B., Kalogera, V. \& Henninger, M. 2004, ApJ,
submitted (astro-ph/0404423)
\end{references}
\end{document}